\documentclass[runningheads,a4paper]{llncs}

\usepackage{amssymb}
\usepackage{graphicx}

\usepackage{url}
\urldef{\mailsa}\path|kokumura@mikan.ap.titech.ac.jp |
\newcommand{\keywords}[1]{\par\addvspace\baselineskip
\noindent\keywordname\enspace\ignorespaces#1}

\begin{document}

\mainmatter  

\title{Analytical Approach to Noise Effects on \\Synchronization in a System of \\Coupled Excitable Elements}

\titlerunning{Analytical Approach to a system of Coupled Excitable Elements}

%
%
\author{Keiji Okumura
\and Masatoshi Shiino}
\authorrunning{Analytical Approach to a system of Coupled Excitable Elements}

\institute{Department of Physics, Faculty of Science, Tokyo Institute of Technology, 2-12-1 Oh-okayama, Meguro-ku, Tokyo 152-8551, Japan\\
\mailsa\\
}

%
%

\toctitle{Lecture Notes in Computer Science}
\tocauthor{Authors' Instructions}
\maketitle

\begin{abstract}
We report relationships between the effects of noise and applied constant currents on the behavior of a system of excitable elements. The analytical approach based on the nonlinear Fokker-Planck equation of a mean-field model allows us to study the effects of noise {\it without approximations} only by dealing with deterministic nonlinear dynamics . We find the similarity, with respect to the occurrence of oscillations involving subcritical Hopf bifurcations, between the systems of an excitable element with applied constant currents and mean-field coupled excitable elements with noise.
\end{abstract}

\section{Introduction}
The effects of noise on neural systems are one of the great concern for researchers in neurosciences. Among theoretical and computational studies is the influence of noise on excitable systems as well as sustained oscillatory ones. Simulations of coupled Hodgkin-Huxley neurons subjected to noise are one of typical examples to demonstrate its behavior \cite{Wang00}. Numerical solutions of the Fokker-Planck equation of coupled active rotators with excitatory and inhibitory elements have been obtained and analysed \cite{Kanamaru03}. Another approach of Gaussian approximations has also been used to reveal bifurcation structures with changes in noise intensity for investigating a system of coupled active rotators and Fitz-Hugh Nagumo neurons \cite{Kurrer95,Zaks05}.

In this paper, we propose a model of coupled excitable elements under the influence of external Langevin noise on the basis of mean-field concept \cite{Shiino01,Ichiki07,Okumura10a}. Satisfying self-averaging property, the model enables us to describe the time evolution of order parameters of the system {\it without approximation}. Investigating this dynamic system instead of a set of Langevin equations, we show similarity between noise and applied currents effects on coupled and uncoupled excitable elements from the viewpoint of bifurcation structures.

\section{Excitable Element}
\subsection{Model}
First, we consider the effects of applied constant currents on a system of an excitable element. Let us suppose that dynamics of a 2-dimensional system of $z^{(x)},z^{(y)}$ satisfies the following set of differential equations:
\begin{eqnarray}
\displaystyle \frac{\textrm{d}z^{(x)}}{\textrm{d}t}&=-a^{(x)}z^{(x)}&+J^{(x)}F^{(x)}(b^{(x,x)}z^{(x)}+b^{(x,y)}z^{(y)})+I\displaystyle, \label{eq:EE1x}\\
\displaystyle \frac{\textrm{d}z^{(y)}}{\textrm{d}t}&=-a^{(y)}z^{(y)}&+J^{(y)}F^{(y)}(b^{(y,x)}z^{(x)}+b^{(y,y)}z^{(y)})\displaystyle, \label{eq:EE1y}
\end{eqnarray}
where $a^{(\mu)},b^{(\mu,\nu)},J^{(\mu)}$ $(\mu=x,y)$ are constants, $I$ is the applied current and $F^{(\mu)}(\cdot)$ are coupling functions. To make the system excitable, we specify $F^{(x)}(\cdot)$ and $F^{(y)}(\cdot)$ as nonlinear and linear functions as
\begin{eqnarray}
&&F^{(x)}(z)=z\displaystyle \exp\left(-\frac{z^{2}}{2}\right),\label{eq:Fx} \\
&&F^{(y)}(z)=z.\label{eq:Fy}
\end{eqnarray}
The model parameter values are $a^{(x)}=2.5$, $a^{(y)}=0.0030$, $b^{(x,x)}=1.5$, $b^{(x,y)}=0.50$, $b^{(y,x)}=4.0$, $b^{(y,y)}=1.0$, $J^{(x)}=5.0$, $J^{(y)}=-0.0040$. In this situation, $z^{(x)}$ and $z^{(y)}$ correspond to fast and slow variables and the system exhibits excitable properties for the applied constant current $I=-3.0$.

\begin{figure}[tb]
 \begin{minipage}{0.5\hsize}
  \begin{center} 
   \includegraphics[scale=0.41,angle=-90,clip]{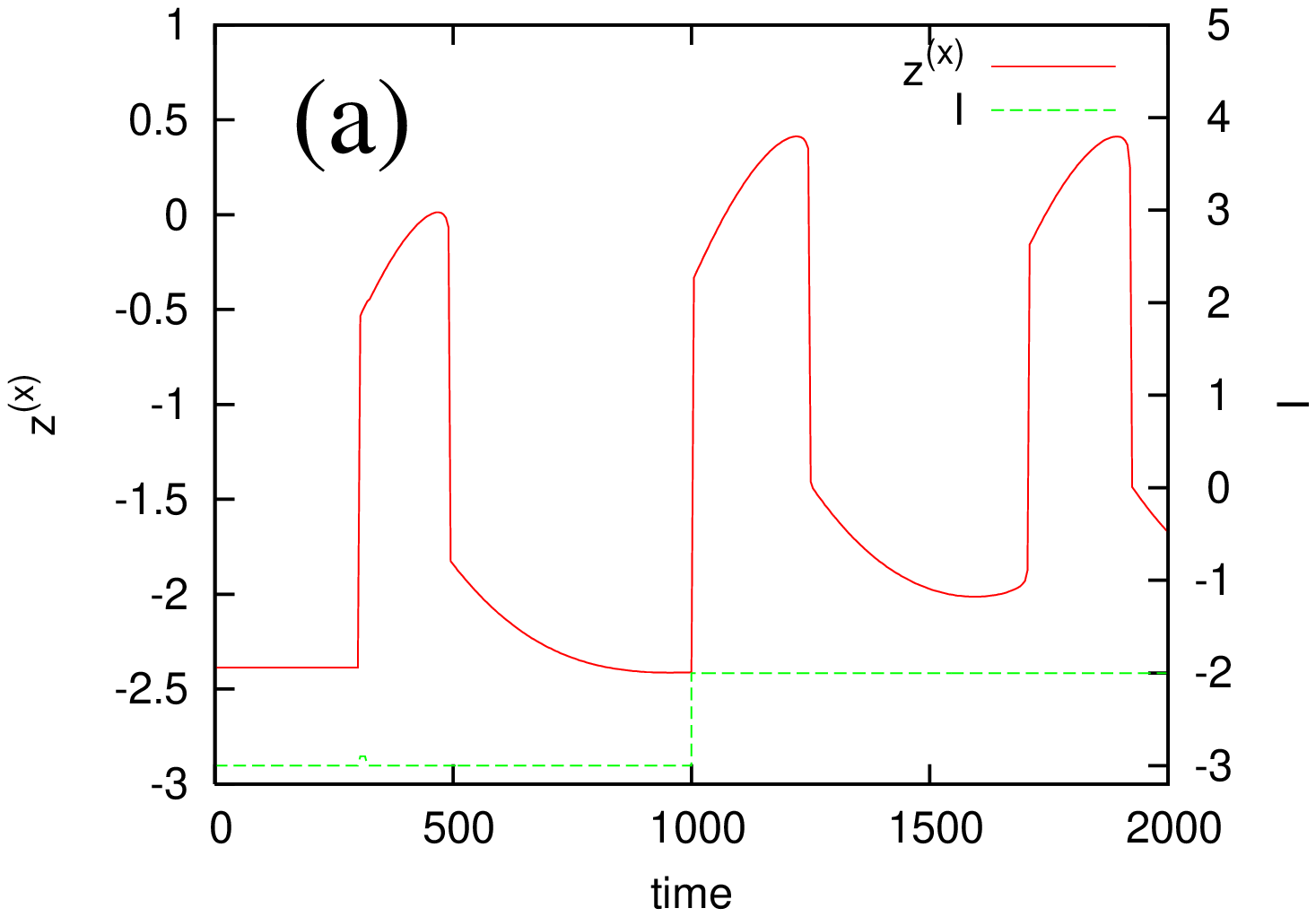}
  \end{center}
 \end{minipage}
 \begin{minipage}{0.5\hsize}
  \begin{center} 
   \includegraphics[scale=0.41,angle=-90,clip]{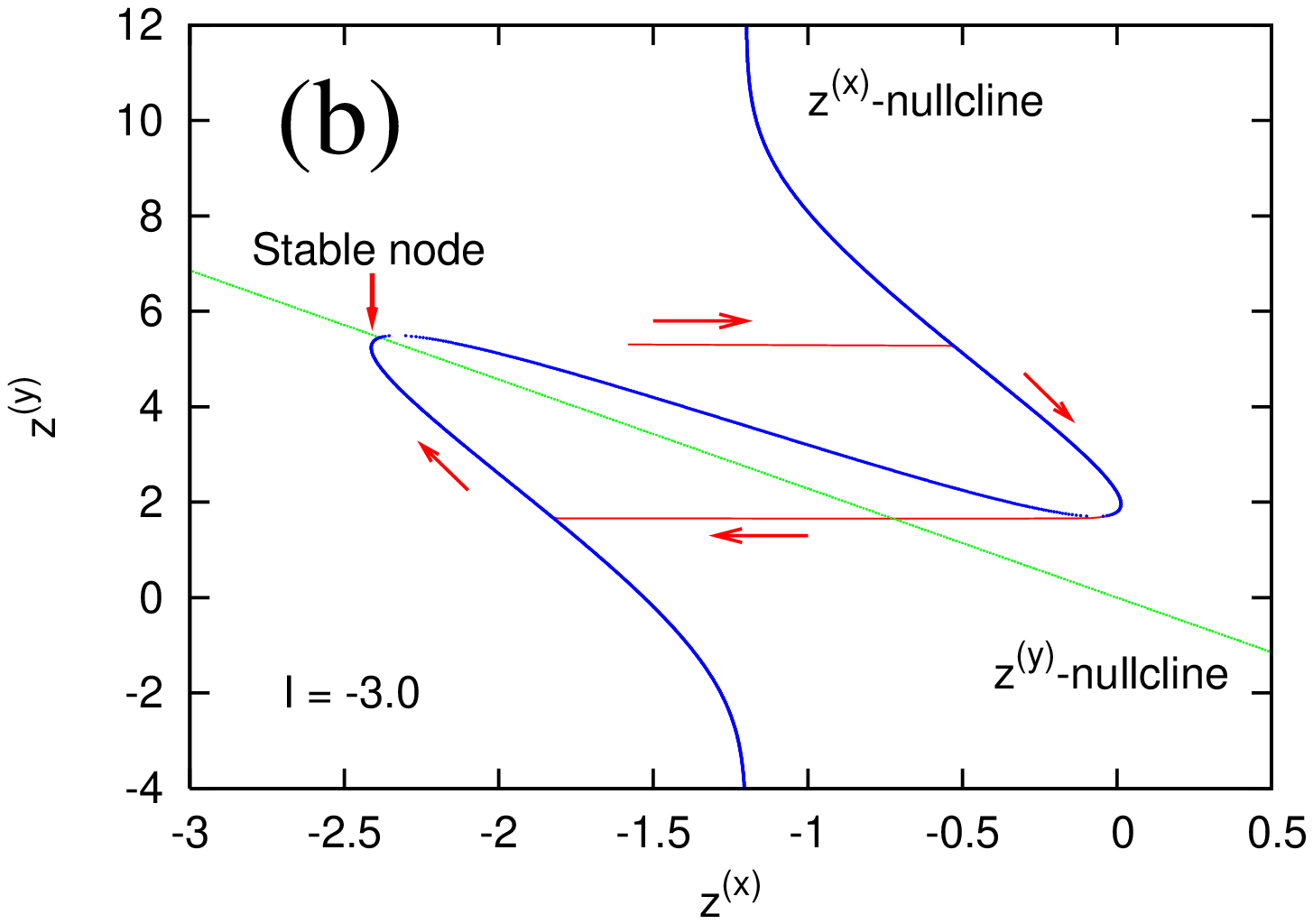}
  \end{center}
 \end{minipage}
 \caption{Dynamics of the system of an excitable element. (a) The time evolution of $z^{(x)}$. $z^{(x)}$ gets excited once by the perturbation at $t=300.0$. After the changes in amplitude of the applied constant current ($t>1000.0$), $z^{(x)}$ begins to oscillate. (b) Nullclines in phase plane. At $I=-3.0$, the trajectory converges to the stable node after sufficiently large times.}\label{fig:EE1}
\end{figure}

Figure \ref{fig:EE1} shows the behavior of the system. When the system is perturbed, $z^{(x)}$ `gets excited' to generate a single pulse, while $z^{(x)}$ exhibits  periodic pulses with a certain level of applied constant currents $I$ (Fig. \ref{fig:EE1} (a)). We can also easily understand this dynamics in phase plane (Fig. \ref{fig:EE1} (b)). Taking an initial condition right side from the equilibrium point, the trajectory begins its journey to return to the fixed point.

\subsection{Bifurcation Diagrams against Applied Constant Currents}
\begin{figure}[tb]
 \begin{minipage}{0.5\hsize}
  \begin{center} 
   \includegraphics[scale=0.41,angle=-90,clip]{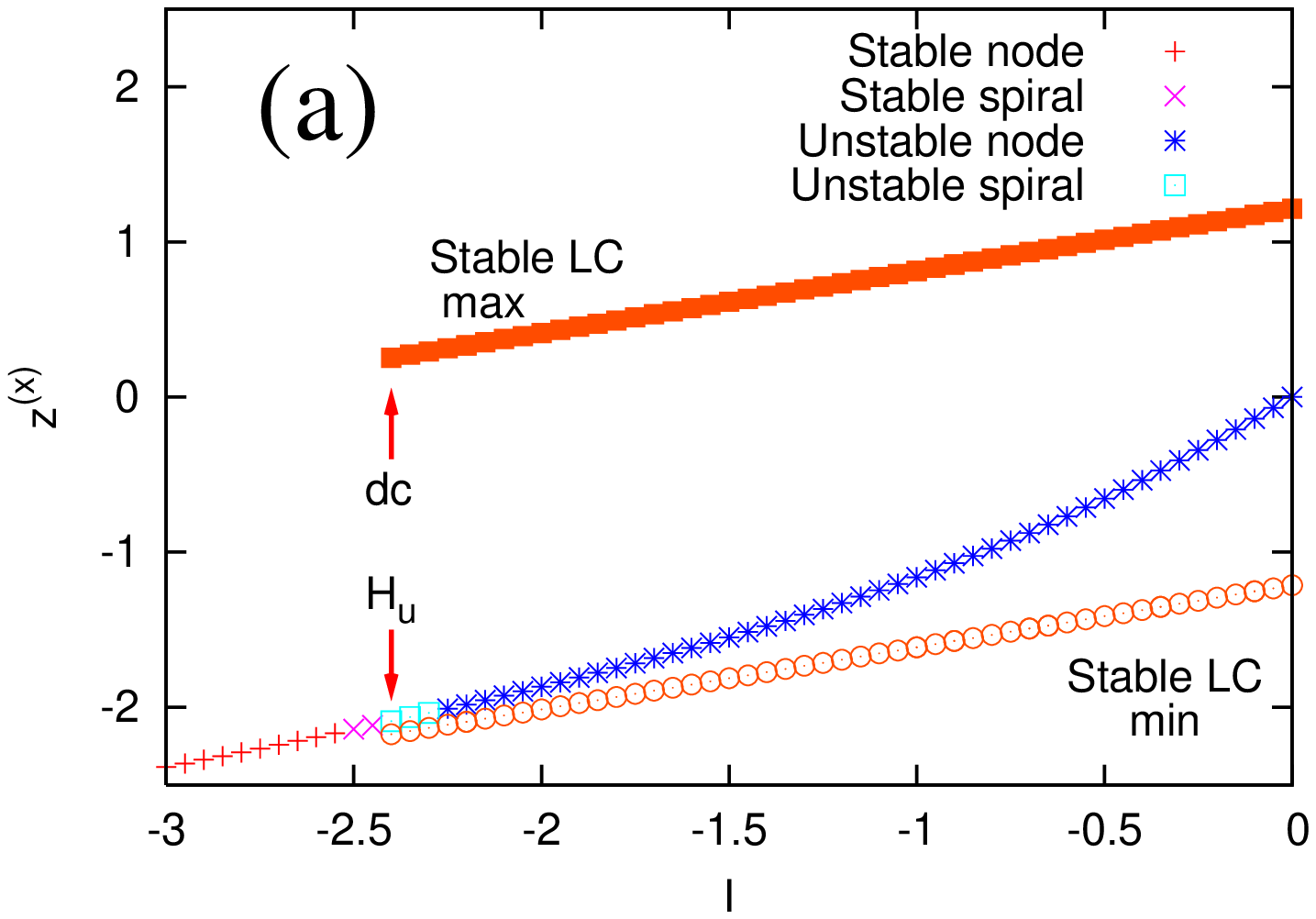}
  \end{center}
 \end{minipage}
 \begin{minipage}{0.5\hsize}
  \begin{center} 
   \includegraphics[scale=0.41,angle=-90,clip]{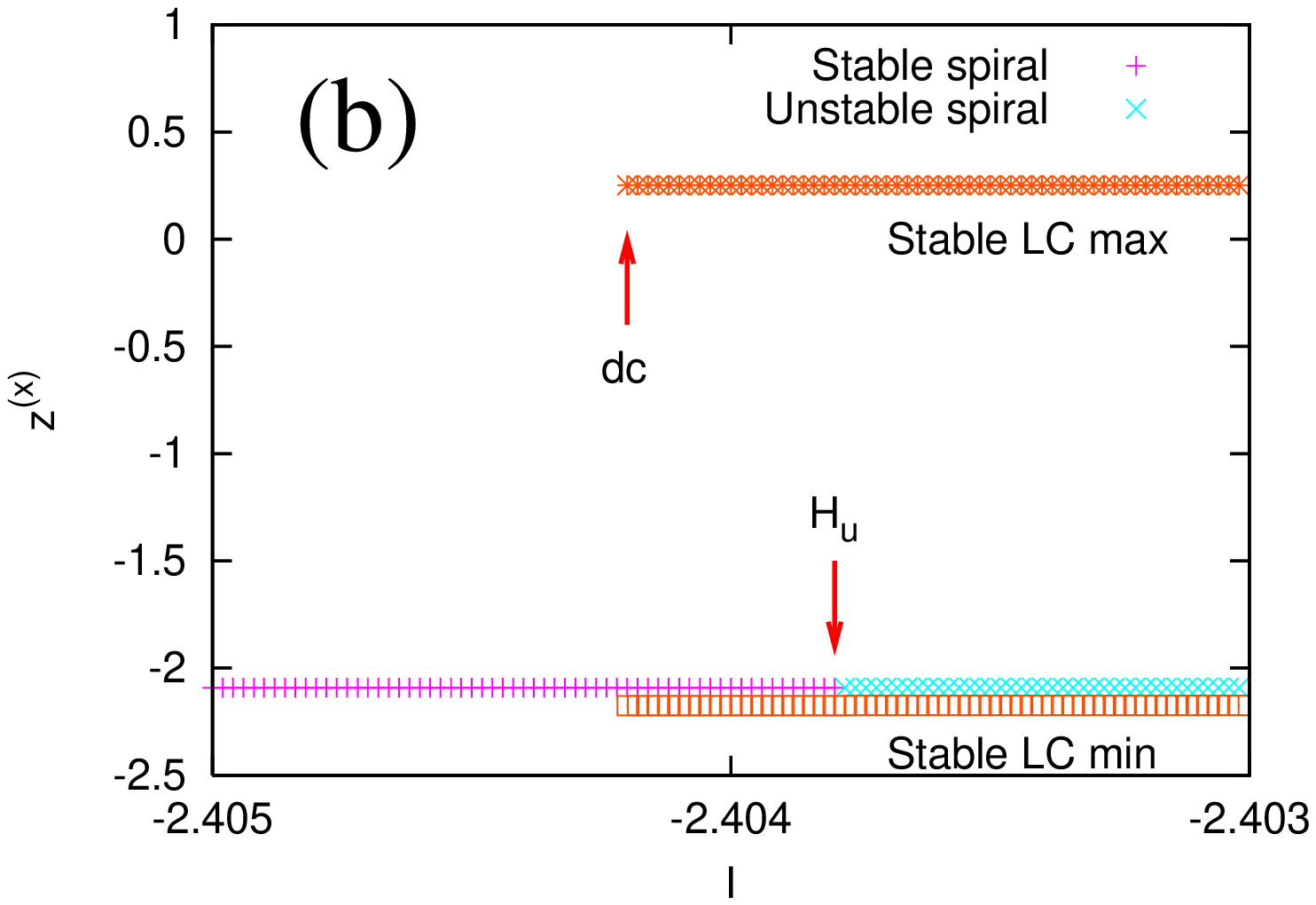}
  \end{center}
 \end{minipage}
 \caption{Bifurcation diagrams of the system of an excitable element. (a) The whole diagram. $H_u$ and $dc$ denote the subcritical Hopf bifurcation and double cycle, respectively. (b) Magnified view of the occurrence of the limit cycle. The limit cycle appears accompanying the subcritical Hopf bifurcation.}\label{fig:EE1_Bf_I}
\end{figure}

To reveal the bifurcation structures with changes in applied constant currents $I$, we conduct a linear stability analysis (Fig. \ref{fig:EE1_Bf_I}). Let us start with the case $I=-3.0$, where the fixed point of the system is the stable node. With increasing the applied constant current $I$, this point moves toward origin and its stability changes to the stable spiral, to the unstable spiral, and to the unstable node, yielding stable limit cycles accompanying the subcritical Hopf bifurcation. Note that the fixed points have rotational symmetry as $(z_0^{(x)},z_0^{(y)},I)$ = $(-z_0^{(x)},-z_0^{(y)},-I)$. In this way, appropriate levels of applied constant currents $|I| < 2.4038 $ give rise to oscillatory states, whereas these oscillations disappear in higher or lower amplitude $|I| > 2.4042 $. In amplitude $2.4038 < |I| < 2.4042$, the limit cycle and fixed point attractors coexist.

\section{Coupled Excitable Elements under the Influence of Noise}
\subsection{Model}
To compare the effects of noise and applied constant currents on excitable elements, we extend the system described by Eqs. (\ref{eq:EE1x}) - (\ref{eq:Fy}) to a coupled one. We suppose a system of N excitable elements and assume that they interact with each other via coupling functions $F^{(\mu)}(\cdot)$ with $J^{(\mu)}$ denoting coupling constants. In addition, each element of a system are subjected to the independent Langevin noise $\eta_{i}^{(\mu)}(t)$. Each of those systems reads
\begin{eqnarray}
\displaystyle \frac{\textrm{d}z_{i}^{(x)}}{\textrm{d}t}&=-a^{(x)}z_{i}^{(x)}&+\displaystyle \frac{1}{N}\sum_{j=1}^{N}J^{(x)}F^{(x)}(b^{(x,x)}z_{j}^{(x)}+b^{(x,y)}z_{j}^{(y)})+I+\eta_{i}^{(x)}(t),\label{eq:CEE1x}\\
\displaystyle \frac{\textrm{d}z_{i}^{(y)}}{\textrm{d}t}&=-a^{(y)}z_{i}^{(y)}&+\displaystyle \frac{1}{N}\sum_{j=1}^{N}J^{(y)}F^{(y)}(b^{(y,x)}z_{j}^{(x)}+b^{(y,y)}z_{j}^{(y)})+\eta_{i}^{(y)}(t), \label{eq:CEE1y}
\end{eqnarray}
where $i=1,\cdots,N$. We postulate that the Langevin noise $\eta_{i}^{(\mu)}(t)$ are white Gaussian ones, $\langle\eta_{i}^{(\mu)}(t)\rangle=0$, $\langle\eta_{i}^{(\mu)}(t)\eta_{j}^{(\nu)}(t^{\prime})\rangle=2D^{(\mu)}\delta_{ij}\delta_{\mu\nu}\delta(t-t^{\prime})$. The coupling functions $F^{(\mu)}(\cdot)$ and the model parameters are the same as described above. The applied constant current is set to $I=-3.0$. In the absence of noise, Eqs. (\ref{eq:CEE1x}) and (\ref{eq:CEE1y}) with $N=1$ recover Eqs. (\ref{eq:EE1x}) and (\ref{eq:EE1y}).

\subsection{Nonlinear Fokker-Planck Equation Approach}
In the thermodynamic limit $N \rightarrow \infty$, we can take advantage of reducing Eqs. (\ref{eq:CEE1x}) and (\ref{eq:CEE1y}) to a single body equation as seen below. The mean-field coupling terms satisfy the self-average property, which is written by the empirical probability density $P(t,z^{(x)},z^{(y)})$,
\begin{equation}
\displaystyle \langle F^{(\mu)}\rangle\equiv\int \textrm{d}z^{(x)}\textrm{d}z^{(y)}F^{(\mu)}(b^{(\mu,x)}z^{(x)}+b^{(\mu,y)}z^{(y)})P(t,z^{(x)},z^{(y)}).\label{eq:MFCoupling1}
\end{equation}
Then the system of Eqs. (\ref{eq:CEE1x}) - (\ref{eq:CEE1y}) is indeed reduced to the 1-body dynamics $z^{(x)}, z^{(y)}$ as
\begin{eqnarray}
\displaystyle \frac{\textrm{d}z^{(x)}}{\textrm{d}t}&=-a^{(x)}z^{(x)}&+J^{(x)}\displaystyle \langle F^{(x)}\rangle+I+\zeta^{(x)}(t),\nonumber\\
\displaystyle \frac{\textrm{d}z^{(y)}}{\textrm{d}t}&=-a^{(y)}z^{(y)}&+J^{(y)}\displaystyle \langle F^{(y)}\rangle+\zeta^{(y)}(t),\nonumber
\end{eqnarray}
with white Gaussian noise $\zeta^{(\mu)}(t)$, $\langle\zeta^{(\mu)}(t)\rangle=0$, $\langle\zeta^{(\mu)}(t)\zeta^{(\nu)}(t^{\prime})\rangle=2D^{(\mu)}\delta_{\mu\nu}\delta(t-t^{\prime})$. Thus, one obtains the nonlinear Fokker-Planck equation for the empirical probability density corresponding to the above Langevin equations \cite{Frank05},
\begin{eqnarray}
&&\displaystyle \frac{\partial}{\partial t}P(t,z^{(x)},z^{(y)})=\nonumber\\
&&-\displaystyle \frac{\partial}{\partial z^{(x)}}\left[-a^{(x)}z^{(x)}+J^{(x)}\langle F^{(x)}\rangle+I-D^{(x)}\frac{\partial}{\partial z^{(x)}}\right]P\nonumber\\
&&-\displaystyle \frac{\partial}{\partial z^{(y)}}\left[-a^{(y)}z^{(y)}+J^{(y)}\langle F^{(y)}\rangle-D^{(y)}\frac{\partial}{\partial z^{(y)}}\right]P.\label{eq:NFPE}
\end{eqnarray}

A Gaussian probability density is a special solution of the nonlinear Fokker-Planck equation (\ref{eq:NFPE}). Furthermore, since the H theorem \cite{Shiino01} ensures that the probability density satisfying Eq. (\ref{eq:NFPE}) converges to the Gaussian-form for sufficiently large times, we are concerned with the Gaussian probability density as

\begin{eqnarray*}
&&P_{\tiny \textrm{G}}(t,z^{(x)},z^{(y)})=
 \displaystyle \frac{1}{2\pi\sqrt{\det C_{\tiny \textrm{G}}(t)}}\exp\left[-\frac{1}{2}\textrm{\boldmath $s$}_{\tiny \textrm{G}}^{T}C_{\tiny \textrm{G}}^{-1}(t)\textrm{\boldmath $s$}_{\tiny \textrm{G}}\right],\nonumber\\
&&\qquad \textrm{\boldmath $s$}_{\tiny \textrm{G}}^{T}=(z^{(x)}-\langle z^{(x)}\rangle_{\tiny \textrm{G}},\,z^{(y)}-\langle z^{(y)}\rangle_{\tiny \textrm{G}})
\equiv(u^{(x)},\,u^{(y)}),\\
&&\qquad C_{{\tiny \textrm{G}} ij}(t)=\langle s_{i}s_{j}\rangle_{\tiny \textrm{G}},
\end{eqnarray*}
where $\langle \cdot \rangle_{\tiny \textrm{G}}$ denotes expectation over $P_{\tiny \textrm{G}}$. Then, the coupling terms Eqs. (\ref{eq:MFCoupling1}) are described only up to the second moments. We derive a set of closed ordinary differential equations as 
\begin{eqnarray}
&&\displaystyle \frac{\textrm{d}\langle z^{(x)}\rangle_{\tiny \textrm{G}}}{\textrm{d}t}=-a^{(x)}\langle z^{(x)}\rangle_{\tiny \textrm{G}}+J^{(x)}\langle F^{(x)}\rangle_{\tiny \textrm{G}}+I, \label{eq:z_meanx} \\
&&\displaystyle \frac{\textrm{d}\langle z^{(y)}\rangle_{\tiny \textrm{G}}}{\textrm{d}t}=-a^{(y)}\langle z^{(y)}\rangle_{\tiny \textrm{G}}+J^{(y)}\langle F^{(y)}\rangle_{\tiny \textrm{G}}, \label{eq:z_meany} \\
&&\displaystyle \frac{\textrm{d}\langle u^{(x)^{2}}\rangle_{\tiny \textrm{G}}}{\textrm{d}t}=-2a^{(x)}\langle u^{(x)^{2}}\rangle_{\tiny \textrm{G}}+2D^{(x)}, \label{eq:z_var} \\
&&\displaystyle \frac{\textrm{d}\langle u^{(y)^{2}}\rangle_{\tiny \textrm{G}}}{\textrm{d}t}=-2a^{(y)}\langle u^{(y)^{2}}\rangle_{\tiny \textrm{G}}+2D^{(y)}, \label{eq:z_var} \\
&&\displaystyle \frac{\textrm{d}\langle u^{(x)}u^{(y)}\rangle_{\tiny \textrm{G}}}{\textrm{d}t}=-(a^{(x)}+a^{(y)})\langle u^{(x)}u^{(y)}\rangle_{\tiny \textrm{G}}, \label{eq:z_cov} 
\end{eqnarray}
where one has from Eqs. (\ref{eq:Fx}) - (\ref{eq:Fy})
\begin{eqnarray}
&&\displaystyle \langle F^{(x)}\rangle_{\tiny \textrm{G}}=\frac{m^{(x)}}{(\sigma^{2}+1)^{3/2}}\exp\left[-\frac{m^{(x)^{2}}}{2(\sigma^{2}+1)}\right],\label{eq:Fx_mean} \\
&&\displaystyle \langle F^{(y)}\rangle_{\tiny \textrm{G}}=m^{(y)},\label{eq:Fy_mean} 
\end{eqnarray}
and $m^{(\mu)}=b^{(\mu,x)}\langle z^{(x)}\rangle_{\tiny \textrm{G}}+b^{(\mu,y)}\langle z^{(y)}\rangle_{\tiny \textrm{G}}$, $\sigma^{2}=b^{(x,x)^{2}}\langle u^{(x)^{2}}\rangle_{\tiny \textrm{G}}+b^{(x,y)^{2}}\langle u^{(y)^{2}}\rangle_{\tiny \textrm{G}}$. Note that $\langle u^{(x)}u^{(y)}\rangle_{\tiny \textrm{G}}\rightarrow 0$ and $\langle u^{(\mu)^{2}}\rangle_{\tiny \textrm{G}}\rightarrow D^{(\mu)}/a^{(\mu)}$ $\, (t\rightarrow\infty)$, implying that the external Langevin noise  contributes to the dynamics through the variance. For simplicity, we assume the Langevin noise intensity $D^{(y)}=0$ in what follows.

Investigating the dynamic system of the order parameter equations written by Eqs. (\ref{eq:z_meanx})-(\ref{eq:z_cov}) together with Eqs. (\ref{eq:Fx_mean}) and (\ref{eq:Fy_mean}), we can compare the effects of noise and applied constant currents on the systems of globally coupled and uncoupled excitable elements.

\begin{figure}[tb]
 \begin{minipage}{0.5\hsize}
  \begin{center} 
   \includegraphics[scale=0.41,angle=-90,clip]{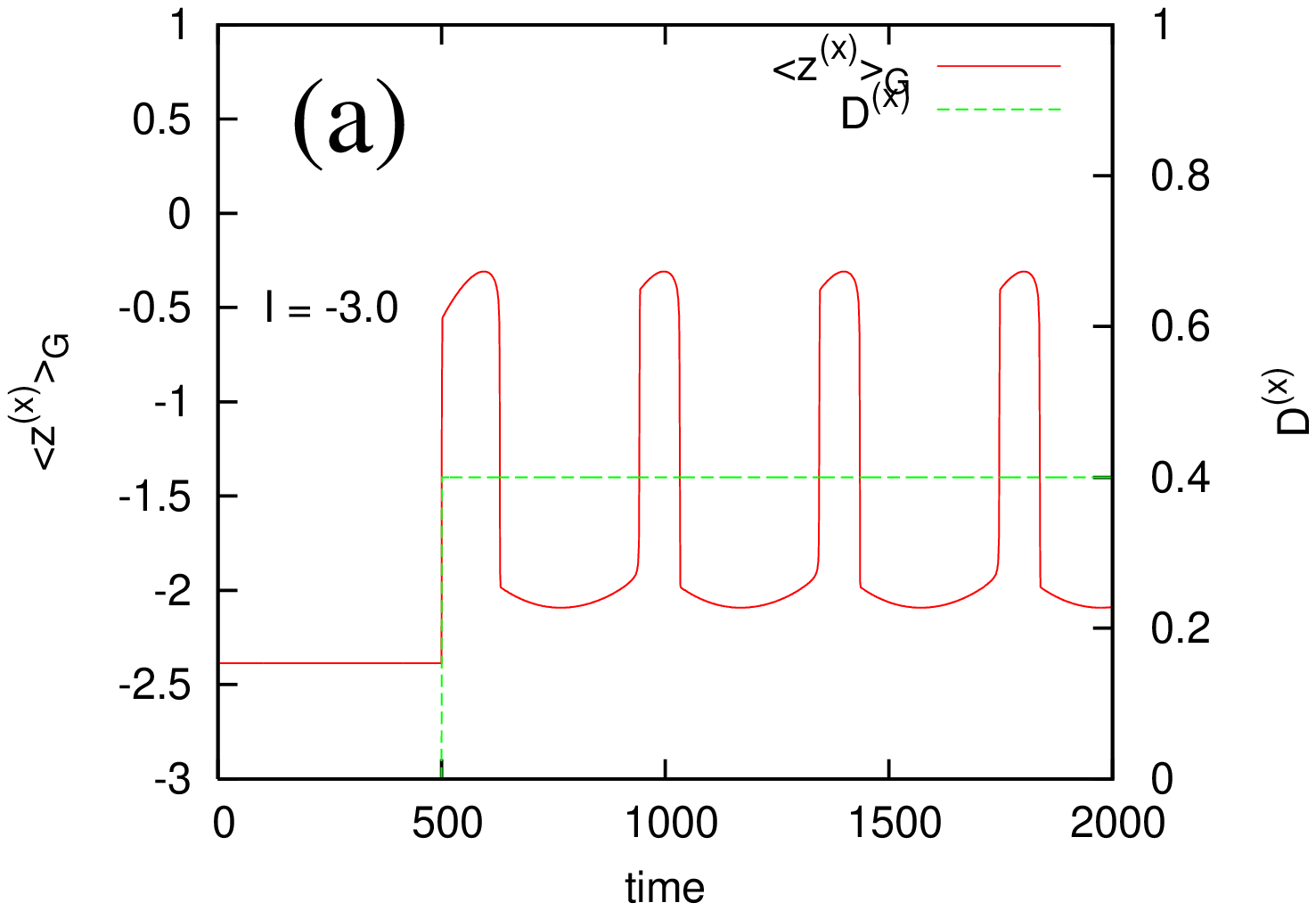}
  \end{center}
 \end{minipage}
 \begin{minipage}{0.5\hsize}
  \begin{center} 
   \includegraphics[scale=0.41,angle=-90,clip]{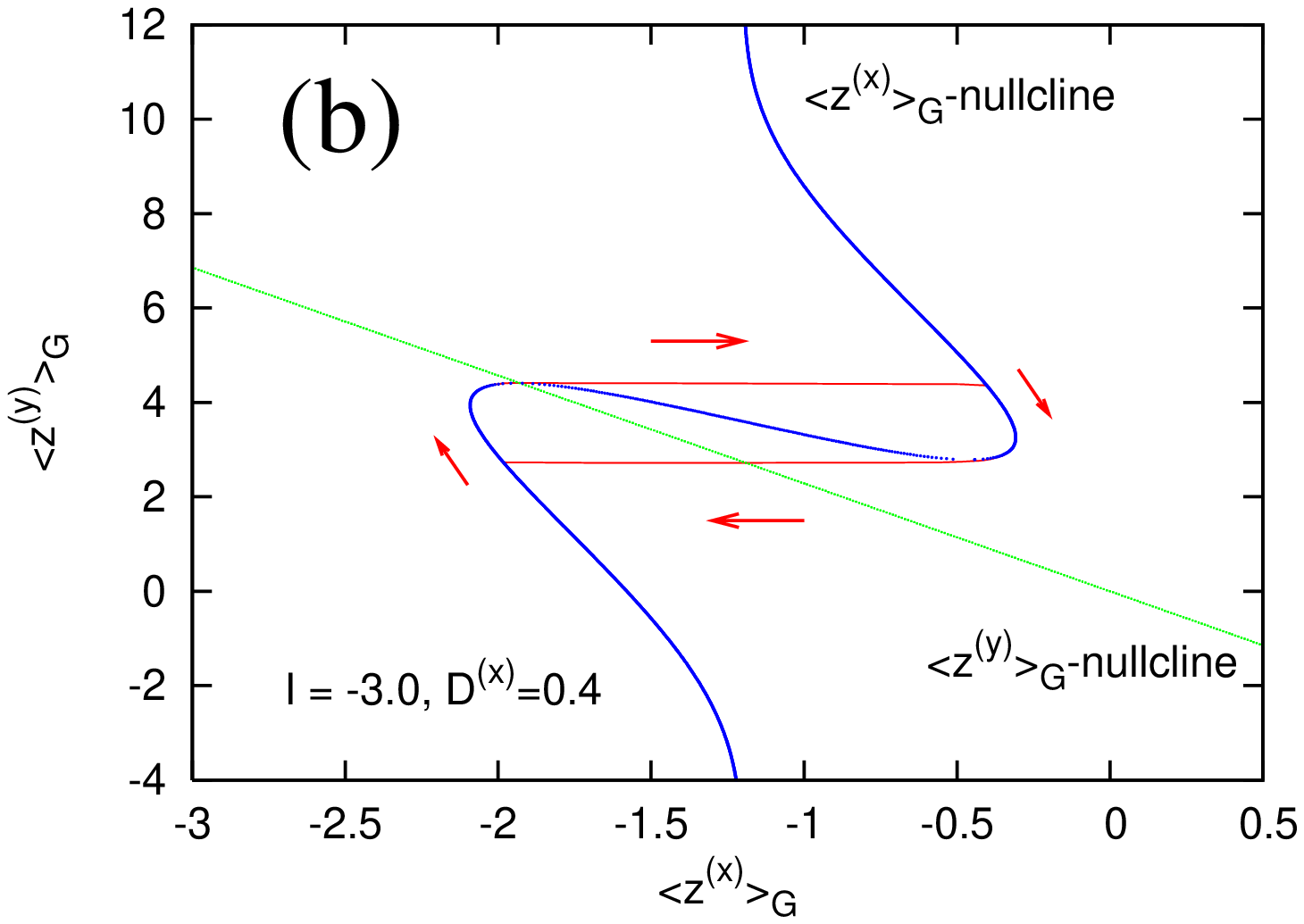}
  \end{center}
 \end{minipage}
 \caption{Dynamics of the system of coupled excitable elements under the influence of noise. (a) The time evolution of $\langle z^{(x)}\rangle_{\tiny \textrm{G}}$. Whereas $z^{(x)}$ does not oscillate for lower noise intensity $D^{(x)}$ ($t<500.0$), it begins to oscillate for an appropriate noise level ($t>500.0$). (b) Nullclines in phase plane with the noise intensity $D^{(x)}=0.4$. In both figures, the applied constant current is set to $I=-3.0$.}\label{fig:CEE1}
\end{figure}

The dynamics of order parameters of the system is shown in Fig. \ref{fig:CEE1}. While the mean value $\langle z^{(x)}\rangle_{\tiny \textrm{G}}$ converges to the stable fixed point for lower intensity of the Langevin noise $D^{(x)}$, it oscillates for relatively higher noise intensity $D^{(x)}$ (Fig. \ref{fig:CEE1} (a)). This oscillatory state suggests that individual excitable elements almost simultaneously get excited under the influence of noise, {\it i.e.} noise-induced synchronization in coupled excitable elements. The portrait of phase space (Fig. \ref{fig:CEE1} (b)) is similar to the case of uncoupled deterministic system (Fig. \ref{fig:EE1} (b)).

\subsection{Nonequilibrium Phase Transitions and Synchronization Induced by Langevin Noise}
\begin{figure}[htb]
  \begin{center} 
    \begin{tabular}{c}
      {\includegraphics[scale=0.41,angle=-90,clip]{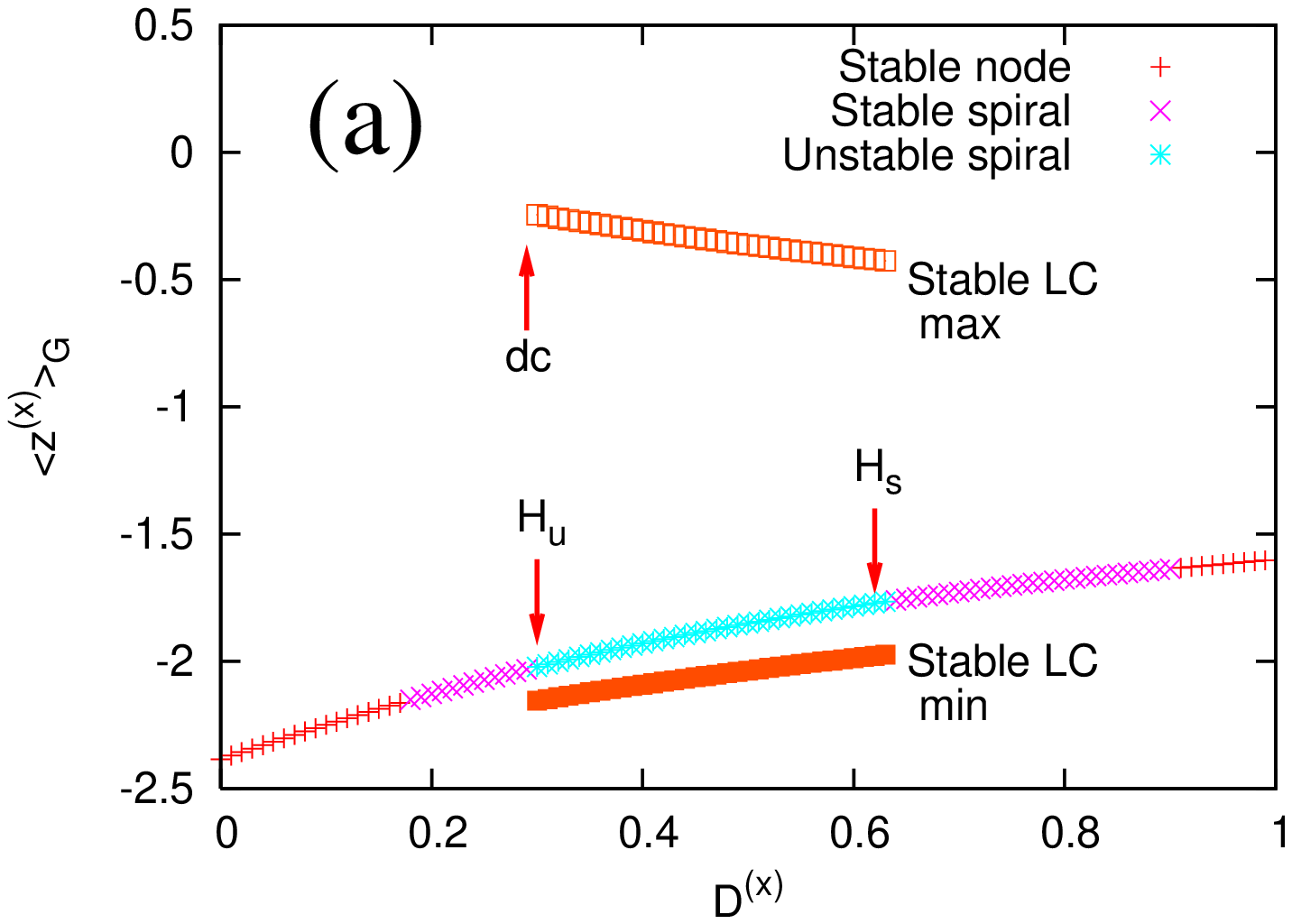}}\\
      \begin{minipage}{0.5\hsize} \begin{center} 
        \includegraphics[scale=0.41,angle=-90,clip]{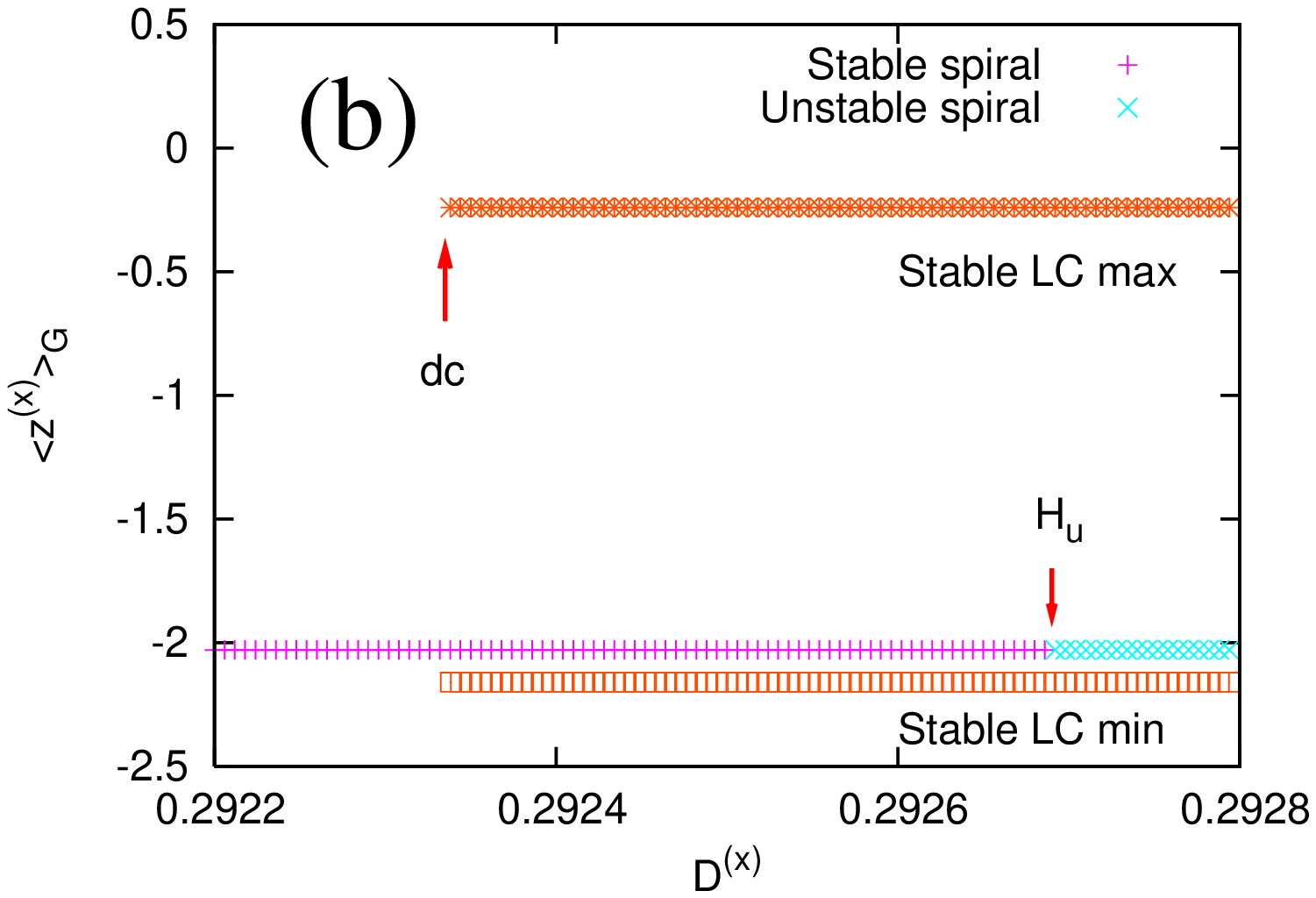}
      \end{center} \end{minipage}
      \begin{minipage}{0.5\hsize} \begin{center} 
        \includegraphics[scale=0.41,angle=-90,clip]{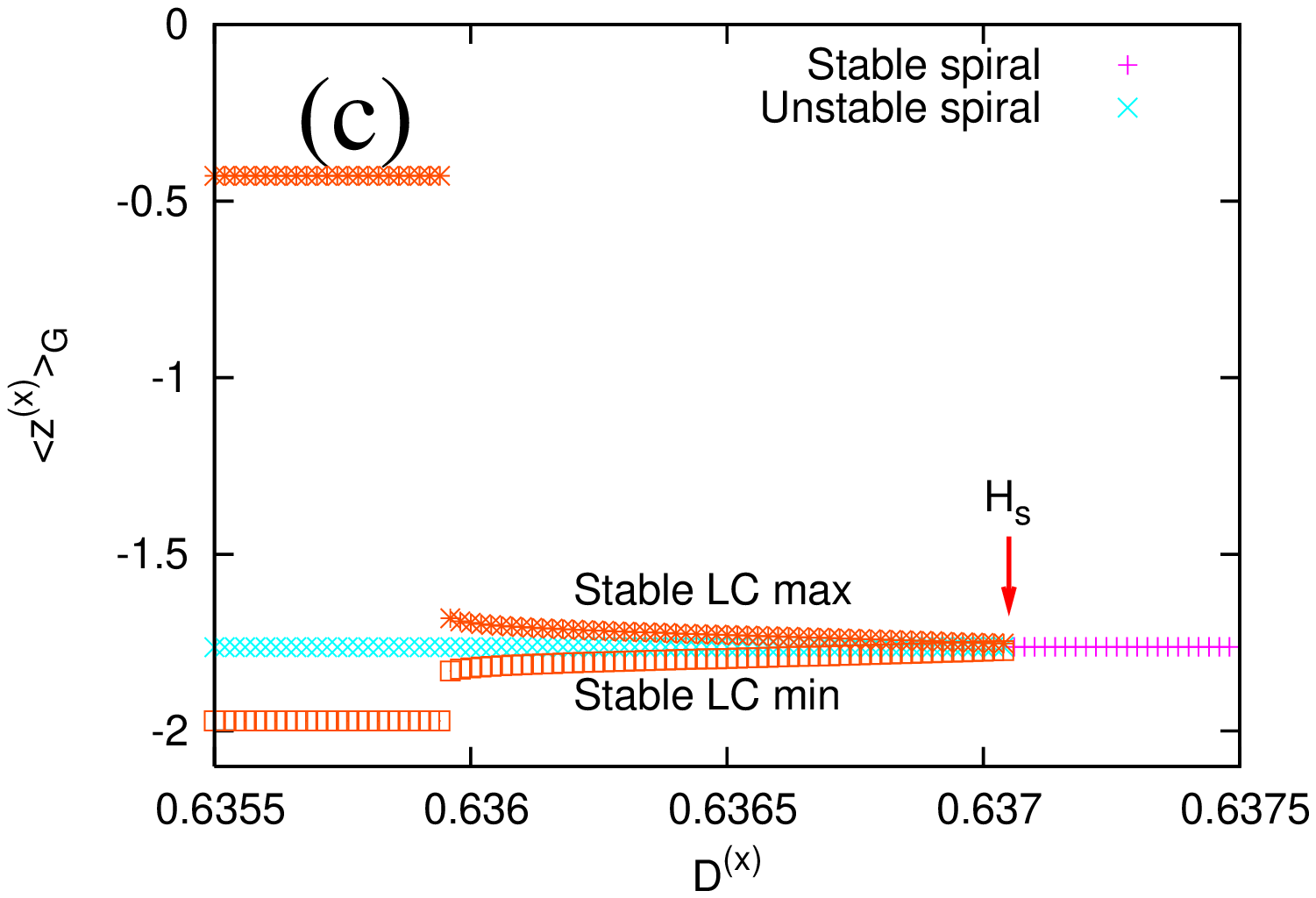}
      \end{center} \end{minipage}
    \end{tabular}
  \end{center}
 \caption{Bifurcation diagrams of the system of coupled excitable elements under the influence of noise. (a) The whole diagram. $H_s$ and $H_u$, and $dc$ denote the supercritical and subcritical Hopf bifurcation, and double cycle, respectively. (b) Magnified view of the occurrence of the limit cycle. The limit cycle appears accompanying the subcritical Hopf bifurcation. (c) Magnified view of  the disappearance of the limit cycle. The limit cycle vanishes with the supercritical Hopf bifurcation.}\label{fig:CEE1_Bf_Dx}
\end{figure}

Using the nonlinear Fokker-Planck equation approach with the mean-field model, we can identify the occurrence of genuine bifurcations of order parameters with changes in noise intensity. The result of numerical survey is shown in Fig. \ref{fig:CEE1_Bf_Dx}. 

In the deterministic limit $D^{(x)} \rightarrow 0$, the fixed point of the system is a stable node. With increasing $D^{(x)}$, this stability changes to a stable spiral and further to an unstable spiral. A limit cycle attractor occurs accompanying the subcritical Hopf bifurcation (Fig. \ref{fig:CEE1_Bf_Dx} (b)), similar to the deterministic uncoupled case with changes in applied constant currents (Fig. \ref{fig:EE1_Bf_I} (b)). In contrast, the disappearance of the oscillatory state is the result of the supercritical Hopf bifurcation (Fig. \ref{fig:CEE1_Bf_Dx} (c)). So it is clearly seen that a certain level of noise induces synchronized oscillations in the system of mean-field coupled excitable elements under the influence of noise.

\section{Summary}
We have studied relationships between the roles of noise and applied constant currents in a system of excitable elements. The analytical approach using the nonlinear Fokker-Planck equation associated with the mean-field model allows us to obtain genuine bifurcation diagrams against noise intensity. In a system of an excitable element without noise, applied constant currents of a certain amplitude have given rise to oscillatory states, which originate from the subcritical Hopf bifurcation. The similar structure has been found in coupled excitable elements under the influence of noise. 

Further analysis of the effects of the multiplicative noise on a system of mean-field coupled excitable elements  will be reported elsewhere.

K.O. acknowledges the financial support from the Global Center of Excellence Program by MEXT, Japan through the ``Nanoscience and Quantum Physics'' Project of the Tokyo Institute of Technology.

%
%
%
\end{document}